# Australia's Approach to AI Governance in Security & Defence

S. Kate Devitt[1,2] & Damian Copeland[1]

[1] University of Queensland

[2] Trusted Autonomous Systems

k.devitt@uq.edu.au ; damian.copeland@uq.edu.au

Acknowledgments: Tim McFarland, Eve Massingham, Rachel Horne & Tara Roberson

## Abstract

*Australia is a leading AI nation with strong allies and partnerships. Australia has prioritised the development of robotics, AI, and autonomous systems to develop sovereign capability for the military. Australia commits to Article 36 reviews of all new means and methods of warfare to ensure weapons and weapons systems are operated within acceptable systems of control. Additionally, Australia has undergone significant reviews of the risks of AI to human rights and within intelligence organisations and has committed to producing ethics guidelines and frameworks in Security and Defence. Australia is committed to OECD's values-based principles for the responsible stewardship of trustworthy AI as well as adopting a set of National AI ethics principles. While Australia has not adopted an AI governance framework specifically for the Australian Defence Organisation (ADO); Defence Science and Technology Group (DSTG) has published 'A Method for Ethical AI in Defence' (MEAID) technical report which includes a framework and pragmatic tools for managing ethical and legal risks for military applications of AI. Australia can play a leadership role by integrating legal and ethical considerations into its ADO AI capability acquisition process. This requires a policy framework that defines its legal and ethical requirements, is informed by Defence industry stakeholders, and provides a practical methodology to integrate legal and ethical risk mitigation strategies into the acquisition process.*

**Introduction**

On 27 February 2021, Australia's Loyal Wingman military aircraft hinted at the possibility of fully autonomous flight at Woomera Range Complex in South Australia (Royal Australian Air Force 2021; Insinna 2021). With no human on board, the plane used a pre-programmed route with remote supervision to undertake and complete its mission. The flight's success and the Royal Australian Air Force's announcement to order six aircraft, signalled an intention to incorporate artificial intelligence (AI) to increase military autonomous capability and freedom of manoeuvre. Air Vice-Marshal (AVM) Cath Roberts (Head of Air Force Capability) said "The Loyal Wingman project is a pathfinder for the integration of autonomous systems and artificial intelligence to create smart human-machine teams" (de Git 2021). AVM Roberts also confirmed that "[w]e need to ensure that ethical and legal issues are resolved at the same pace that the technology is developed" (Department of Defence 2021b).

Just over six months later, on 25th September, Australia[1] won the silver medal at the 2021 DARPA Subterranean Challenge, also known as the *robot Olympics*. In the event, Australia used multiple robotics platforms equipped with AI to autonomously explore, map, and discover models representing lost or injured people, suspicious backpacks, or phones, or navigate tough conditions such as pockets of gas. The outstanding performance confirms Australia's

---

[1] Team included Commonwealth Scientific and Industrial Research Organisation (CSIRO) Data61 and Emesent; plus International partner Georgia Institute of Technology

international reputation at the forefront of robotics, autonomous systems and AI research and development (Persley 2021).

In parallel to technology development, Australia is navigating the challenge of developing and promoting AI governance structures inclusive of Australian values, standards, ethical and legal frameworks. National initiatives have been led by:

1. Australia's national research organisation, CSIRO Data61 (Hajkowicz et al. 2019),
2. Government (Department of Industry Innovation and Science 2019; Department of Industry Science Energy and Resources 2021, 2020a, 2020b) in the civilian domain; and
3. Defence Science and Technology Group, Royal Australian Air Force and Trusted Autonomous Systems in Defence (Devitt et al. 2021; Department of Defence 2021b).

Two large surveys of Australian attitudes to AI were conducted in 2020. Selwyn and Gallo Cordoba (2021) found that, based on over 2000 respondents, Australian public attitudes towards AI are informed by an educated awareness of the technologies affordances. Attitudes are generally positive and respondents are proud of their scientific achievements. However, Australians are concerned about the government's trustworthiness using automated decision-making algorithms due to incidents such as the Robodebt (Braithwaite 2020) and #CensusFail (Galloway 2017) scandals. A second study of over 2500 respondents found that Australians have low trust in AI systems but generally 'accept' or 'tolerate' AI (Lockey, Gillespie, and Curtis 2020). They found that Australians trust research institutions and Defence organisations the most to use AI and trusted commercial organisations the least. Australians expect AI to be regulated and carefully managed (Lockey, Gillespie, and Curtis 2020).

This chapter will begin with Australia's strategic position, Australia's definition of AI and identifying what ADO wants AI for. It will then move into AI governance initiatives and specific efforts to develop frameworks for ethical AI both in both civilian and military contexts. The chapter will conclude with likely future directions for Australia in AI governance to reshape military affairs and Australia's role in international governance mechanisms and strategic partnerships.

**Australia's strategic position**

Australia's current strategic position has been shaped by two major developments, 1) the announcement of AUKUS 15 September 2021 (Morrison, Johnson, and Biden 2021) and 2) public commitment to NATO and AUKUS allies through sanctions, supply of humanitarian support and lethal weapons to the Ukraine to defend against the invasion of Russia (Prime Minister and Minister of Defence 2022, 1 March 2022).

AUKUS confirmed a shift in strategic interests for the United States to Asia Pacific and away from the Middle East with the withdrawal from Afghanistan.

> "Recognizing our deep defense ties, built over decades, today we also embark on further trilateral collaboration under AUKUS to enhance our joint capabilities and interoperability. These initial efforts will focus on cyber capabilities, artificial intelligence, quantum technologies, and additional undersea capabilities" (The White House 2021b)

Commitment to defend Ukraine signals a strengthening global alliance amongst liberal democracies, of which Australia is a part. The Prime Minister calls the conflict ""a very big wake-up call" that reunites liberal democracies to meet a polarising and escalating threat" (Kelly 2022).

The 2020 Strategic Update identified new objectives for Australian defence—see Box 1.

**Box 1 New Objectives for Australian Defence 2020** (Department of Defence (2020a, 24-25)

1. to **shape** Australia's strategic environment;
2. to **deter** actions against Australia's interests; and
3. to **respond** with credible military force, when required (Department of Defence 2020a 2.12).

These new objectives will guide all aspects of ADO's planning including force structure planning, force generation, international engagement and operations (2.13).

To implement these objectives, ADO will:
- prioritise our immediate region (the north-eastern Indian Ocean, through maritime and mainland South East Asia to Papua New Guinea and the South West Pacific) for the ADF's geographical focus;
- grow the ADF's self-reliance for delivering deterrent effects; expand Defence's capability to respond to grey-zone activities, working closely with other arms of Government;
- enhance the lethality of the ADF for the sorts of high-intensity operations that are the most likely and highest priority in relation to Australia's security; maintain the ADF's ability to deploy forces globally where the Government chooses to do so, including in the context of US-led coalitions; and
- enhance ADO's capacity to support civil authorities in response to natural disasters and crises.

With these in mind, Australia's global position has been elevated with the announcement of a new Australia, UK and United States science and technology, industry, and defence partnership (AUKUS) (Morrison, Johnson, and Biden 2021). This partnership is likely to increase data, information and AI sharing and aligned AI governance structures and interoperability policies (Deloitte Center for Government Insights 2021) to manage joint and cooperative military action, deterrence, cyber-attacks, data theft, disinformation, foreign interference, economic coercion, attacks on critical infrastructure, supply chain disruption and so forth (Hanson and Cave 2021).

In addition to AUKUS, Australia has a number of strategic partnerships including the global 'five-eyes' network of UK, US, Australia, Canada and New Zealand (Office of the Director of National Intelligence); the Quad surrounding China India, Japan, Australia and US (Shih and Gearan 2021) and local partnerships including the Association of Southeast Asian Nations (ASEAN) (Thi Ha 2021) and Pacific family (Blades 2021).

The strategic issues identified in both the Defence White Paper (Department of Defence 2016) and the 2020 Strategic Update (Department of Defence 2020a) puts AI among priority information and communications technology capabilities, e.g.

> "3.39 Over the next five years, Defence will need to plan for developments including next generation secure wireless networks, artificial intelligence, and augmented analytics" (Department of Defence 2020a).

**Australia's definition of artificial intelligence**

Australia (Department of Industry Science Energy and Resources 2021) defines AI as:

> "AI is a collection of interrelated technologies that can be used to solve problems autonomously and perform tasks to achieve defined objectives. In some cases, it can do this without explicit guidance from a human being (Hajkowicz et al. 2019). AI is more than just the mathematical algorithms that enable a computer to learn from text, images or sounds. It is the ability for a computational system to sense its environment, learn, predict and take independent action to control virtual or physical infrastructure."

Australia defines AI by its functions (sensing, learning, predicting, independent action), focus and degree of independence in the achievement of defined objectives with or without explicit guidance from a human being. The Australian definition encompasses the role of AI in digital and physical environments without discussing any particular methodology or technology that might be used. In doing so it aligns itself with the OECD's Council on Artificial Intelligence's definition of an AI system:

> "An AI system is a machine-based system that can, for a given set of human-defined objectives, make predictions, recommendations, or decisions influencing real or virtual environments. AI systems are designed to operate with varying levels of autonomy" (OECD Council on Artificial Intelligence 2019)

What does Australian Defence want artificial intelligence for?

While the Australian Department of Defence in Australia has not formally adopted a Defence AI Roadmap or Strategy, Australia has prioritised developing sovereign AI capabilities—see Box 2—as well as: robotics, autonomous systems, precision guided munitions, hypersonic weapons, and integrated air and missile defence systems; space; and information warfare and cyber capabilities (Australian Government 2021).

**Box 2 Australian Sovereign Industry Capability Priority:**

**Robotics, Artificial Intelligence and Autonomous Systems** (Australian Government (2021)

Robotics and autonomous systems are an important element of military capability. They act as a force multiplier and protect military personnel and assets.

The importance of these capabilities will continue to grow over time. Robotics and autonomous systems will become more prevalent commercially and in the battlespace.

Australian industry must have the ability to design and deliver robotic and autonomous systems. This will enhance the ADF's combat and training capability through:
- improving efficiency
- reducing the physical and cognitive load to the operator
- increasing mass
- achieving decision making superiority
- decreasing risk to personnel.

These systems will comprise of:
- advanced robots
- sensing and artificial intelligence encompassing al gorithms
- machine learning and deep learning.

> These systems will enhance bulk data analysis. This will facilitate decision making processes and enable autonomous systems.

Australia notes that AI will play a vital role in ADO's future operating environment, delivering on strategic objectives of shape, deter and respond (Department of Defence 2021a). AI will contribute to Australia in maintaining a capable, agile and potent ADO.

Potential military AI applications have been taxonomized into warfighting functions (force application, force protection, force sustainment, situational understanding) and enterprise functions (personnel, enterprise logistics, business process improvement)—see Annex A (Devitt et al. 2021). Operational contexts help discern the range of purposes of AI within Defence as well as diverse legal, regulatory, and ethical structures required in each domain to govern AI use. For example the use of AI in ensuring abidance with workplace health and safety risk management (Morrison 2021; Centre for Work Health and Safety 2021) is relevant to Defence People Group[2], whereas the use of AI within new weapons systems and the Article 36 review process is within the portfolio for Defence Legal (Commonwealth of Australia 2018). AI will also be needed to manage Australia's grey zone threat (Townshend, Lonergan, and Warden 2021).

The ADO acknowledges that they need to effectively use their data holdings to harness the opportunities of AI technologies and the Defence Artificial Intelligence Centre (DAIC) has been established to accelerate Defence's AI capability (Department of Defence 2021a, 35; 2020b). The ADO has launched *The AI for Decision Making Initiative* (Defence Science Institute 2020a, 2021; Defence Science & Technology Group 2021a) and a Defence Artificial

---

[2] See https://www1.defence.gov.au/about/people-group

Intelligence Research Network (DAIRNET) to develop AI "to process noisy and dynamic data in order to produce outcomes to provide decision superiority" (Defence Science & Technology Group 2021b).

These efforts include some human-centred projects, such as *human factors for explainable AI* and studies in AI bias in facial recognition (Defence Science Institute 2020b). Defence has committed to develop guidelines on the ethical use of data (Department of Defence 2021a) and the Australian Government has committed to governance and ethical frameworks for the use of artificial capabilities for intelligence purposes (Attorney-General's Department 2020, Recommendation 154). Ethics guidelines will help Australia respond to public debate on the ethics of facial recognition for military purposes even if biases are reduced (van Noorden 2020).

**AI Governance in Australia**

AI governance includes social, legal, ethical and technical layer (algorithms and data) that require norms, regulation, legislation, criteria, principles, data governance, algorithm accountability and standards (Gasser and Almeida 2017). Australia's strategic, economic, cultural, diplomatic, and military use of AI will be expected to be governed in accordance with Australian attitudes and values (such as described in Box 3) and international frameworks.

---

**Box 3 Australian Values (Department of Home Affairs 2020)**

- Respect for the freedom and dignity of the individual
- Freedom of religion (including the freedom not to follow a particular religion), freedom of speech and freedom of association

---

> - Commitment to the rule of law, which means that all people are subject to the law and should obey it
> - Parliamentary democracy whereby our laws are determined by parliaments elected by the people, those laws being paramount and overriding any other inconsistent religious or secular 'laws'
> - Equality of opportunity for all people, regardless of their gender, sexual orientation, age, disability, race or national or ethnic origin
> - A 'fair go' for all that embraces:
>   - mutual respect;
>   - tolerance;
>   - compassion for those in need;
>   - equality of opportunity for all
> - The English language as the national language, and as an important unifying element of Australian society

Australia is positioning itself to be consistent with emerging best practice internationally for a for ethical, trustworthy (Ministère des Armées 2019), responsible AI (Fisher 2020) and allied frameworks for ethical AI in Defence (Lopez 2020; Stanley-Lockman 2021).

To this end, Australia is a founding member of The Global Partnership on AI (GPAI)[4], an international and multi-stakeholder initiative to undertake cutting-edge research and pilot projects on AI priorities to advance the responsible development and use of AI built around a shared commitment to the *OECD Recommendation on Artificial Intelligence*[5]. The OECD has demonstrated considerable "ability to influence global AI governance through epistemic authority, convening power, and norm- and agenda-setting" (Schmitt 2021).

---

[4] Other GPAI countries include: Canada, France, Germany, India, Italy, Japan, Mexico, New Zealand, Korea, Singapore, Slovenia, the United Kingdom, the United States, and the European Union. In December 2020, Brazil, the Netherlands, Poland, and Spain joined GPAI (Gobal Partnership on AI 2021; Department of Industry Science Energy and Resources 2020a).

Since 2018, Australia has used a consultative methodology and public communication of evidence-based ethics frameworks in both civil and military domains. The civil domain work driven by CSIRO's Data61 (Dawson et al. 2019) and the military work driven by Defence Science and Technology Group (DSTG) (Devitt et al. 2021).

**AI Ethics Principles**

In 2019, the Australian Government sought public submissions in response to a CSIRO Data61AI Ethics discussion paper (Dawson et al. 2019). A voluntary AI Ethics Framework emerged from the Department of Industry Innovation and Science (2019) (DISER) to guide businesses and governments developing and implementing AI in Australia. The framework includes eight AI ethics principles (AU-EP) to help reduce the risk of negative impacts from AI and ensure the use of AI is underpinned by good governance standards—see Box 4.

---

[5] GPAI working groups are focussed on four key themes: responsible AI; data governance; the future of work; and innovation and commercialisation

> **Box 4 Australia's AI Ethics Principles (AU-EP)** (Department of Industry Innovation and Science (2019)
>
> 1. **Human, societal, and environmental wellbeing**: AI systems should benefit individuals, society, and the environment.
> 2. **Human-centred values**: AI systems should respect human rights, diversity, and the autonomy of individuals.
> 3. **Fairness**: AI systems should be inclusive and accessible and should not involve or result in unfair discrimination against individuals, communities, or groups.
> 4. **Privacy protection and security**: AI systems should respect and uphold privacy rights and data protection and ensure the security of data.
> 5. **Reliability and safety**: AI systems should reliably operate in accordance with their intended purpose.
> 6. **Transparency and explainability**: There should be transparency and responsible disclosure so people can understand when they are being significantly impacted by AI and can find out when an AI system is engaging with them.
> 7. **Contestability**: When an AI system significantly impacts a person, community, group or environment, there should be a timely process to allow people to challenge the use or outcomes of the AI system.
> 8. **Accountability**: People responsible for the different phases of the AI system lifecycle should be identifiable and accountable for the outcomes of the AI systems, and human oversight of AI systems should be enabled.

Case studies have been undertaken with industry to evaluate the usefulness and effectiveness of the principles. Many of the findings and due diligence frameworks will be useful in the dialogue between Defence industries, Australian Defence Force and the Department of Defence.

Key findings from Industry (Department of Industry Science Energy and Resources 2020b) include:

- AU-EP are relevant to any organisation involved in AI (private, public, large or small)
- Organisations expect the Australian Government to lead by example and implement AU-EP.
- Implementing AU-EP can ensure that businesses can exemplify best practice and be ready to meet community expectations or any changes in standards or laws
- Ethical issues can be complex, and businesses may need more help from professional or industry bodies, academia or experts and government.
- Businesses need training and education, certification, case study examples, and cost-effective methods to help them implement and utilise AU-EP.

The case studies revealed that responsibilities of AI purchasers and AI developers differ. Each group needed internal due diligence, communication and information from external stakeholders, including vendors or customers, to establish accountability and responsibility obligations. Businesses found some principles more challenging to practically implement. The advice given by the Government is that businesses ought to document the process of managing ethical risks (despite ambiguity) and to refer serious issues to relevant leaders.

To support ethical AI businesses are advised by DISER to:

- Set appropriate standards and expectations of responsible behaviour when staff deploy AI. For example, via a responsible AI policy and supporting guidance.

- Include AI applications in risk assessment processes and data governance arrangements.

- Ask AI vendors questions about the AI they have developed.

- Form multi-disciplinary teams to develop and deploy AI systems. They can consider and identify impacts from diverse perspectives.

- Establish processes to ensure there is clear human accountability for AI-enabled decisions and appropriate senior approvals to manage ethical risks. For example, a cross-functional body to approve an AI system's ethical robustness.

- Increase ethical AI awareness raising activities and training for staff.

The Australian Government commits to continuing to work with agencies to encourage greater uptake and consistency with AU-EP (Department of Industry Science Energy and Resources 2020b). In the 2021 AI Action Plan, Australia hopes that widespread adoption of AU-EP among business, government and academia will build trust in AI systems (Department of Industry Science Energy and Resources 2021). AU-EP are included in the Defence Method for Ethical AI in Defence report (Appendix A Comparison of Ethical AI Frameworks Devitt et al. 2021, 48-50). However, some have questioned the value of AU-EP without them being embedded in policy, practise and accountability mechanisms. Contentious uses of AI in government such as facial recognition by police have not gone through ethical review and do not have institutional ethical oversight (ASPI 2022). While federal frameworks may not have seen effective operationalisation, NSW has published their own AI Assurance Framework that mandates AI projects to undergo ethical risk assessment (NSW Government 2022).

**Standards**

The Future of Humanity Institute at the University of Oxford recommends developing international standards for ethical AI research and development (Cihon 2019; Dafoe 2018). This is consistent with Standards Australia's *Artificial Intelligence Standards Roadmap: Making Australia's Voice Heard* (2020)—see Box 5.

Standards Australia seeks to increase cooperation with the United States National Institute for Standards & Technology (NIST) and other Standards Development Organisations (SDOs). Australia has a stated aim to participate in ISO/IEC/JTC 1/SC 42, and the National Mirror Committee (IT-043) regarding AI. Standards Australia notes the importance of improving AI data quality as well as ensuring Australia's adherence to both domestic and international best practise in privacy and security by design.

**Box 5 Artificial Intelligence Standards Roadmap: Making Australia's Voice Heard**

(Standards Australia (2020)

*Recommendations*:

1. Increase the membership of the Artificial Intelligence Standards Mirror Committee in Australia to include participation from more sectors of the economy and society.

2. Explore avenues for enhanced cooperation with the United States National Institute for Standards & Technology (NIST) and other Standards Development Organisations (SDOs) with the aim of improving Australia's knowledge and influence in international AI Standards development.

3. The Australian Government nominate government experts to participate in ISO/IEC/JTC 1/SC 42, and the National Mirror Committee (IT-043). The Australian Government should also fund and support their participation, particularly at international decision-making meetings where key decisions are made, within existing budgetary means.

4. Australian businesses and government agencies develop a proposal for a direct text adoption of ISO/IEC 27701 (Privacy Information Management), with an annex mapped to local Australian Privacy Law requirements. This will provide Australian businesses and the community with improved privacy risk management frameworks that align with local requirements and potentially those of the GDPR, CBPR and other regional privacy frameworks.

5. Australian Government stakeholders, with industry input, develop a proposal to improve data quality in government services, to optimise decision-making, minimise bias and error, and improve citizen interactions.

> 6. Australian stakeholders channel their concerns about inclusion, through participating in the Standards Australia AI Committee (IT-043), to actively shape the development of an international management system Standard for AI as a pathway to certification.
>
> 7. The Australian Government consider supporting the development of a security-by-design initiative, which leverages existing standards used in the market, and which recognises and supports the work being carried out by Australia's safety by-design initiative.
>
> 8. Develop a proposal for a Standards hub setup to improve collaboration between standards-setters, industry certification bodies, and industry participants, to trial new more agile approaches to AI Standards for Australia.

**Human Rights**

A report by the Australian Human Rights Commissioner (AHRC) (Santow 2021) concerning human rights and AI in Australia makes a suite of recommendations including the establishment of an AI Safety Commissioner (Sadler 2021). Of relevance to this chapter are the recommendations to:

- require human rights impact assessments (HRIA) before any government department or agency uses an AI-informed decision-making system to make administrative decisions [Recommendation 2]

- the government needs to make AI decision making transparent and explainable to affected individuals and give them recourse to challenge the decision [Recommendations 3, 5, 6, 8].

- AU-EP should be used to encourage corporations and other non-government bodies to undertake a human rights impact assessment before using an AI-informed decision-making system.

Human rights impact assessments of AI in Defence will vary depending on the context of deployment, e.g. whether the AI is deployed within the war fighting or rear echelon functions. It is notable that Defence has established precedence working collaboratively with the Australian Human Rights Commission (AHRC), e.g. on 'Collaboration for Cultural Reform in Defence' examining human rights issues including gender, race and diversity, sexual orientation and gender identity and the impact of alcohol and social media on the cultural reform process (Jenkins 2014). Thus, it is possible that Australia could work again with the AHRC on AI decision making in the Australian Defence Force. As algorithms could unfairly bias recommendations, honours and awards, promotion, duties, or postings against particular groups e.g. women or LGBTIQ+. While not documented in the military yet, Tech giants including Amazon have withdrawn AI powered recruitment when they found that it was biased against women (Parikh 2021). There are also new methods to debias the development, use and iteration of AI tools in human resources (Parikh 2021).

**Australian AI Action Plan**

In the Australian AI Action Plan (Department of Industry Science Energy and Resources 2021), the Australian government commits to:

- Developing and adopting AI to transform Australian businesses
- Creating an environment to grow and attract the world's best AI talent
- Using cutting edge AI technologies to solve Australia's national challenges
- Making Australia a global leader in responsible and inclusive AI

To achieve the final point, Australia commits to AI Ethics Principles (Department of Industry Innovation and Science 2019) and the OECD Principles (2019) on AI to promote AI that is innovative, trustworthy and that respects human rights and democratic values.

The OECD AI principles (2019) are:

1. AI should benefit people and the planet by driving inclusive growth, sustainable development and well-being.
2. AI systems should be designed in a way that respects the rule of law, human rights, democratic values and diversity, and they should include appropriate safeguards – for example, enabling human intervention where necessary – to ensure a fair and just society.
3. There should be transparency and responsible disclosure around AI systems to ensure that people understand AI-based outcomes and can challenge them.
4. AI systems must function in a robust, secure and safe way throughout their life cycles and potential risks should be continually assessed and managed.
5. Organisations and individuals developing, deploying or operating AI systems should be held accountable for their proper functioning in line with the above principles.

In 2021, the OECD released a report on how Nations are responding to AI Ethics principles. In the report, Australia is noted as:

- deploying a myriad of policy initiatives, including: establishing formal education programmes on STEM and AI-related fields to empower people with the skills for AI and prepare for a fair labour market transition.

- offering fellowships, postgraduate loans, and scholarships to increase domestic AI research capability and expertise and retain AI talent. Australia has dedicated AUD 1.4 million to AI and Machine Learning PhD scholarships
- Australia and Singapore, building on their pre-existing trade agreement, also signed the Singapore-Australia Digital Economy Agreement (SADEA) where Parties agreed to advance their co-operation on AI

Recently the US and Europe confirmed commitment to OECD principles in a joint statement that:

> "The United States and European Union will develop and implement AI systems that are innovative and trustworthy and that respect universal human rights and shared democratic values, explore cooperation on AI technologies designed to enhance privacy protections, and undertake an economic study examining the impact of AI on the future of our workforces" (The White House 2021a).

Australia is likely to remain aligned with the AI frameworks of allies, particularly the UK (AI Council 2021) and the USA (National Security Commission on Artificial Intelligence 2021).

**AI governance in Defence**

While Australia has not released an overarching AI governance framework for Defence, this chapter outlines an argument for such a framework that draws from publicly released concepts, strategy, doctrine, guidelines, papers, reports and methods relating to human, AI, and data governance relevant to Defence and Australia's strategic position.

Australia is a founding partner in the US's AI Partnership for Defense (PfD) that includes Canada, Denmark, Estonia, France, Finland, Germany, Israel, Japan, the Republic of Korea, Norway, the Netherlands, Singapore, Sweden, the United Kingdom, and the United States (JAIC Public Affairs 2021, 2020). In doing so, Australia has aligned its AI partnerships with AUKUS, five-eyes (minus New Zealand), the Quad (minus India) and ASEAN via Singapore[6]. In particular Australia is seeking to increase AI collaboration with the US and UK through AUKUS (Nicholson 2021).

**AI in Weapons Systems**

Computer software designed to perform computational or control functions has been used in weapons systems for over 40 years (Department of Defense 1978). Such weapons require thorough test and evaluation to identify and mitigate risks of computer malfunction. This has lead to a recent drive for digital engineering (88th Air Base Wing Public Affairs 2019; National Security Commission on Artificial Intelligence 2021). AI and autonomous weapons system (AWS) do not necessarily coincide, but the application of Australia's international and domestic legal obligations to AI weapon systems will almost certainly affect Australia's ability to develop, acquire and operate autonomous military systems. Australia has stated that it considers a sweeping prohibition of AWS to be premature (Australian Permanent Mission and Consulate-General Geneva 2017; Commonwealth of Australia 2018; Senate Foreign Affairs Defence and Trade Legislation Committee 2019, 65) and emphasises the importance in compliance with the legal obligation to undertake Article 36 reviews to manage the legal risks associated with these systems.

---

[6] Note there is little representation from remaining ASEAN nations Brunei, Cambodia, Indonesia, Laos, Malaysia, Myanmar, the Philippines, Thailand and Vietnam or Pacific Nations

Article 36 of the Protocol Additional to the Geneva Conventions of 12 August 1949, and relating to the Protection of Victims of International Armed Conflicts, 8 June 1977 (Additional Protocol 1), provides:

> "In the study, development, acquisition or adoption of a new weapon, means or method of warfare, a High Contracting Party is under an obligation to determine whether its employment would, in some or all circumstances, be prohibited by this Protocol or by any other rule of international law applicable to the High Contracting Party."

The Article 36 process requires Australia to determine whether it can meet its international legal obligations in operating AWS. Performing a thorough Article 36 review requires consideration of International Humanitarian Law ('IHL') prohibitions and restrictions on weapons, including Customary International Law, and an analysis of the normal or expected use of the AWS against the IHL rules governing the lawful use of weapons (i.e. distinction, proportionality and precautions in attack). This includes ensuring weapon operators understand their functions and limitations as well as the likely consequences of their use. Thus, users of AWS are legally required to be reasonably confident about how they will operate before deploying them (Liivoja et al. 2020).

The ADF Concept for Future robotics and autonomous systems (Vine 2020) states:

> "3.10 Existing international law covers the development, acquisition and deployment of any new and emerging capability, including future autonomous weapons systems."

> "3.44 Australia has submitted two working papers to the LAWS GGE in an attempt to demonstrate how existing international humanitarian law is sufficient to regulate current and envisaged weapon systems; the first (Commonwealth of Australia 2018) explained the article 36 weapon review process and the second (Australian Government 2019) outlined the 'System of Control' which regulates the use of force by the ADF. Within the domestic legal system, the RAS (particularly drones) is being considered in the development and review of legislation on privacy, intelligence services and community safety."

Australia argues that "if states uphold existing international law obligations…there is no need to implement a specific ban on AWS, at this time" (Commonwealth of Australia 2019).

However, the 2015 Senate Committee Report on unmanned platforms said 'the committee is not convinced that the use of AWS should be solely governed by the law of armed conflict, international humanitarian law and existing arms control agreements. A distinct arms control regime for AWS may be required in the future" (see para 8.30). The report recommended that:

> "8.33 … the Australian Government support international efforts to establish a regulatory regime for autonomous weapons systems, including those associated with unmanned platforms."

Australia welcomes discussion (e.g. McFarland 2021, 2020) around international legal frameworks on autonomous weapons and how technological advances in weapons systems can comply with international humanitarian law (Senate Foreign Affairs Defence and Trade Legislation Committee 2019).

**Ethical AI Statements Across the Services**

Different defence institutions in Australia have addressed the importance of ethical and legal aspects of AI in their operations. The Royal Australian Navy (2020) stated that "development of trusted autonomous systems is expected to increase accuracy, maintain compliance with Navy's legal and policy obligations as well as regulatory standards, and if utilised during armed conflict, minimise incidental harm to civilians". The Army (2018) said it would "remain cognisant of the ethical, moral and legal issues around the use of RAS technologies as this strategy evolves and is implemented". Finally, the Royal Air Force (RAAF) (2019, 10-11) mentioned that it would explore ways to ensure ethical and moral values and legal accountabilities remain central, including continuously evaluating which decisions can be made by machines and which must be made by humans. The exploration and pursuit of augmented intelligence must be transparent and accountable to the RAAF's legal, ethical, and moral values and obligations. Greater engagement with risk and opportunity must be matched by accountability and transparency.

It is assumed that the governance of AI will dovetail with aspects of human governance, particularly where AI augments or replaces human decision-makers, and in some parts similarly to technology governance and in accordance with best practice in data-governance. Australian Defence has confirmed commitments to non-AI governance of humans and technology as detailed in the section below.

**Human Governance in Defence**

Expectations of human decision-makers are likely to be applied if not extended whenever AI influences or replaces human decision-making, including moral and legal responsibilities. Australian definition of Command ADDP 00.1 Command and Control AL1 (Department of Defence 2019, 1-1):

> Command: The authority which a commander in the military Service lawfully exercises over subordinates by virtue of rank or assignment.
>
> Notes:

> 1. Command includes the authority and responsibility for effectively using available resources and for planning the employment of organising, directing, coordinating and controlling military forces for the accomplishment of assigned missions.
>
> 2. It also includes responsibility for health, welfare, morale and discipline of assigned personnel.

Within the Definition of Command is authority and responsibility over military decision making including the use of physical or digital resources such as how and when AI is deployed. When making decisions, the Australian Defence Force Leadership Doctrine (ADF-P-0, Ed. 3)(2021a) states "Ethical leadership is the single most important factor in ensuring the legitimacy of our operations and the support of the Australian people".

Suggesting that Command is expected to deploy digital assets ethically, the Leadership Doctrine argues in no uncertain terms that "your responsibility as a leader is to ensure the pursuit of your goals is ethical and lawful. There are no exceptions" (Australian Defence Force 2021a, 7). The Australian Defence Force – Philosophical – 0 Military Ethics Doctrine (Australian Defence Force 2021b) breaks down ethical leadership into a framework including intent, values, evaluate, lawful and reflect (see Figure 1).

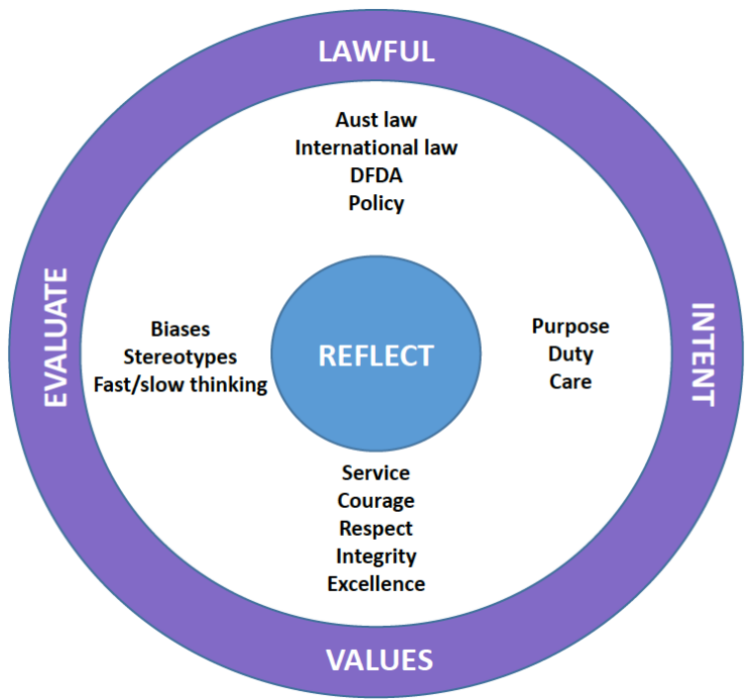

*Figure 1:* Australian Defence Force Ethical Decision-Making Framework (Figure 5.1, Australian Defence Force 2021b).

The *Lead the Way: Defence Transformation Strategy* articulates that Defence wants human decision-makers to be agile, adaptive and ethical with a continuous improvement culture "embedding strong Defence values and behaviours, clear accountabilities and informed and evidence-based decision-making" (Department of Defence 2020c, 21). ADF-P-7 The Education Doctrine (Australian Defence Doctrine Publication 2021) emphasises the importance of "Innovative and inquiring minds" that are "better equipped to adapt to fast-changing technological, tactical and strategic environments". Abilities sought include:

- objectively seek and identify credible information,
- accurately recognise cues and relationships,
- quickly make sense of information and respond appropriately.

Ethical AI could contribute and augment human capabilities for a faster and more agile force.

Australian Defence personnel using AI to augment decision making will be expected to use it ethically and lawfully; to increase the informativeness and evidence-base for decisions; and for decision-makers to agile and accountable both with and without AI.

**Ethics in Australian Cybersecurity and Intelligence**

Key strategic threats for Australia are cybercrime, ransomware and information warfare. Cyber security incidents are increasing in frequency, scale and sophistication, threatening Australia's economic prosperity and national interests (White 2021). AI is likely to play a role in both decision support and in autonomous defence and offensive campaigns to thwart those who seek to undermine Australia's interests.

Australia has not published an ethics of AI policy for cybersecurity or intelligence. However, ethical behaviours are highlighted in publicly available value statements, such as "we always act legally and ethically" (Australian Signals Directorate 2019a, 2019b) and communications suggestive that Australia would expect strong governance of AI systems used in these operations including abidance with domestic and international law and the values of government organisations. For example, speaking to the Lowy Institute Director-General of Australian Signals Directorate (ASD), Mike Burgess (2019) highlighted that "rules guide us when people are watching; values guide us when they're not" (Lowy Institute 2019 51:46) and that ASD is "an organisation that is actually incredibly focused on doing the right thing by the public and being lawful that's an excellent part of our culture born out of our values we put a lot of effort focusing on that" (Lowy Institute 2019 52:27).

A comprehensive review of the legal framework of the national intelligence community highlights the importance of accountability, transparency and oversight of how the Australian government collects and uses data (Richardson 2020).

The government response to the Richardson report (Attorney-General's Department 2020, 40-41) agrees that governance and ethical frameworks should be developed for the use of artificial capabilities for intelligence purposes (recommendation 154), citing values including control, oversight, transparency, and accountability (recommendations 155-156). The Australian government noted the importance of human-in-the-loop decision-making where a person's rights or interests may be affected or where an agency makes an adverse decision in relation to a person (recommendation 155).

The Australian Government is also committed to working with businesses on potential legislative changes including the role of privacy, consumer and data protection laws (Cyber Digital and Technology Policy Division 2020).

**Defence Data Strategy**

In 2021, Defence released a Defence Data Strategy. The strategy promises a Data Security Policy to ensure the adoption of a risk-based approach to data security that allows Defence more latitude to respond to the increase in grey-zone activities, including cyber-attacks, and foreign interference, and a renewed focus on data security and storage processes. Defence identified ethical considerations as a key component of their data strategy—see Box 6. While they commit to being informed by The Australian Code for the Responsible Conduct of Research and the National Statement on Ethical Conduct in Research (Australian Research

Council 2020), neither of these codes provides any guidance on the development of AI for Defence or security purposes.

---

**Box 6 Ethical data, Defence Data Strategy 2021-2023** (Department of Defence 2021a, 42)

Guidelines around the ethical use of data will be developed to ensure we have a shared understanding of our legislative and ethical responsibilities

…

The Australian Code for the Responsible Conduct of Research and the National Statement on Ethical Conduct in Research will inform these guidelines. The ethical use of data guidelines will form part of the Defence Human and Animal Research Manual and policies.

---

Nevertheless, Defence is committed to producing training so that personnel are "equipped to treat data securely and ethically" by 2023 (Department of Defence 2021a, 13).

**Framework for Ethical AI in Defence**

Australia has not adopted an ethics framework specifically for AI use in Defence. However, a Defence Science and Technology technical report based on outcomes from an evidence-based workshop[7] has recommended a method for ethical AI in Defence (MEAID) (Department of Defence 2021b; Devitt et al. 2021) and an Australia-specific framework to guide ethical risk mitigation. MEAID draws from the workshop for further consideration and does not represent

---

[7] Workshop held in Canberra 30 Jul to 1 Aug 2019 with 104 people from 45 organisations including representatives from Defence, other Australian government agencies, the Trusted Autonomous Systems Defence Cooperative Research Centre (TASDCRC), civil society, universities and Defence industry

the views of the Australian Government. Rather than stipulating principles, MEAID identifies five facets of ethical AI and corresponding questions to support science and technical considerations for the potential development of Defence policy, doctrine, research and project management: **Responsibility** – who is responsible for AI?; **Governance** – how is AI controlled?; **Trust** – how can AI be trusted?; **Law** – how can AI be used lawfully? And **Traceability** – How are the actions of AI recorded?—see Figure 2.

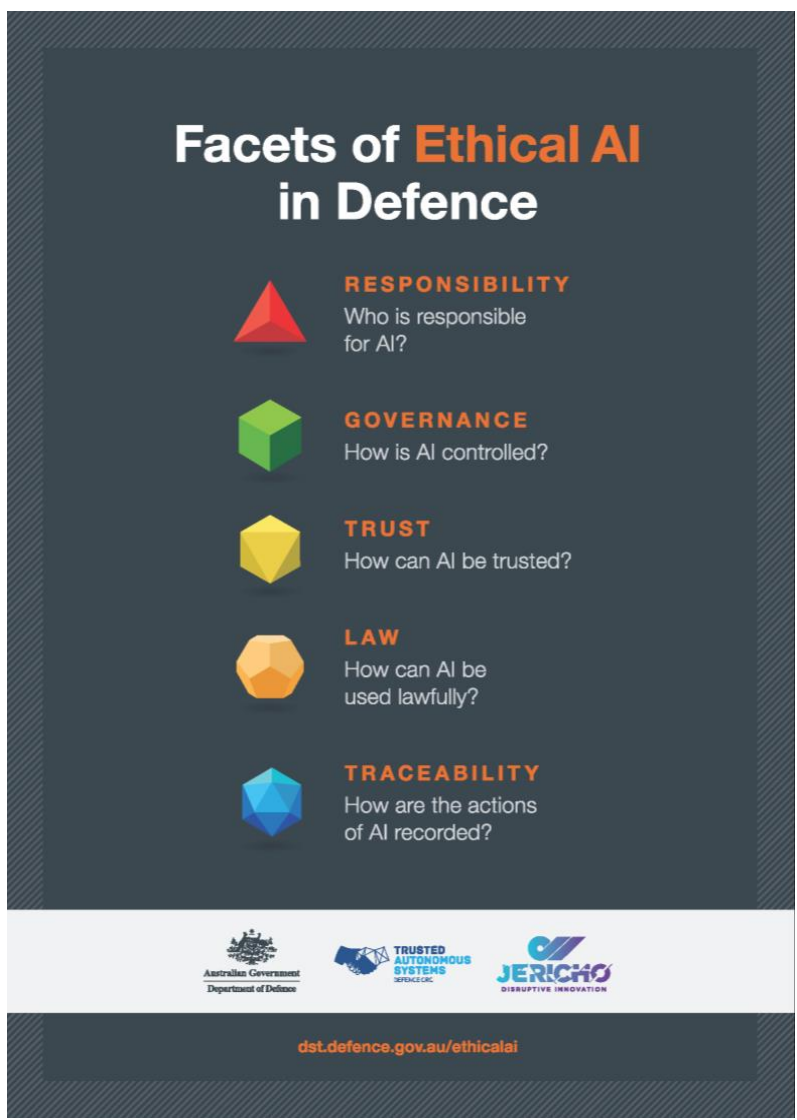

*Figure 2:* Facets of Ethical AI in Defence

MEAID notes that facets of ethical AI for Defence and the associated questions align with the unique concerns and regulatory regimes to which Defence is subject. For example, in times of conflict, Defence is required to comply with international humanitarian law (IHL, lex specialis) and international human rights law (lex generalis) in armed conflict (jus in bello[8]). Defence is also required to comply with international legal norms with respect to the use of force when not engaged in armed conflict (jus ad bellum) when applying military force. Australia's inclusion of 'Law' as an ethical facet highlights the values Australia promotes through abidance with international humanitarian law, particularly the concepts of proportionality, distinction and military necessity which have no direct non-military equivalent and as such requires consideration of a specific set of requirements and responsibilities.

*Responsibility: Who is responsible for AI?*

MEAID notes two key challenges of understanding and responsibility that must be addressed when operating with AI systems. Firstly, in order to effectively and ethically employ a given system (AI or not), the framework argues that a commander must sufficiently understand its behaviour and the potential consequences of its operation (Devitt et al. 2021, 11). Secondly, there can be difficulty in identifying any specific individual responsible for a given decision or action.

Responsibility for critical decisions is spread across multiple decision-makers offering multiple opportunities to exercise authority but also to make mistakes. The allocation of ethical and legal responsibility could be distributed across the nodes/agents in the human-AI network causally relevant for a decision (Floridi 2016). However, legal responsibility ultimately lies with humans. Additionally, AI could help reduce mistakes and augment human makers who bear responsibility (Ekelhof 2018). Decisions made with the assistance of or by AI are captured by accountability frameworks including domestic and international law[9].

---

[8] 'jus in bello' usually refers to specifically to IHL even though human rights law still operates in conflict

The Department of Defence can examine legal cases of responsibility in the civilian domain to guide some aspects of the relevant frameworks, e.g. the apportioning of responsibility for the test-driver in an Uber automated vehicle accident (Ormsby 2019). Defence could also consider arguments that humans within complex systems without proactive frameworks risk being caught in *moral crumple zones* (Elish 2019) where the locus of responsibility falls on human operators rather than the broader system of control within which they operate. Defence must keep front of mind that humans, not AI, have legal responsibilities, and that Individuals—not only states—can bear criminal responsibility directly under international law (Cryer, Robinson, and Vasiliev 2019).

*Governance – how is AI controlled?*

MEAID suggests that AI creators must consider the context in which AI is to be used and how AI will be controlled. The point of interface through which control is achieved will vary, depending on the nature of the system and the operational environment. There must be work conducted to understand how humans can be capable of operating ethically within machine-based systems of control in accordance with Australia's commitment to Article 36 reviews of all new means and methods of warfare (Commonwealth of Australia 2018)[10].

---

[9] There is sometimes considerable uncertainty about exactly how to apply legal frameworks to decisions made with significant AI involvement.

[10] . The Protocol Additional to the Geneva Conventions of 12 August 1949, and relating to the Protection of Victims of International Armed Conflicts (Protocol I), 8 June 1977 refers alternately to ''methods or means of warfare'' (Art. 35(1) and (3), Art. 51(5)(a), Art. 55(1)), ''methods and means of warfare'' (titles of Part III and of Section I of Part III), ''means and methods of attack'' (Art. 57(2)(a)(ii)), and ''weapon, means or method of warfare'' (Art. 36) (International Committee of the Red Cross 2006)

With regards to the control of lethal autonomous weapons, Australia notes the legal, policy, technical, and professional forms of controls imposed systematically throughout the 'life' cycle of weapons over nine stages:

System of control of weapons (Commonwealth of Australia 2019):

> Stage One: Legal and Policy Framework
>
> Stage Two: Design and Development
>
> Stage Three: Testing, Evaluation and Review
>
> Stage Four: Acceptance, Training and Certification
>
> Stage Five: Pre-deployment Selection
>
> Stage Six: Weapon Use Parameters
>
> Stage Seven: Pre-deployment Certification and Training
>
> Stage Eight: Strategic and Military Controls for the Use of Force
>
> Stage Nine: After-Action Evaluation

MEAID supports the governance framework of IEEE's Ethically Aligned Design by the IEEE Global Initiative on Ethics of Autonomous and Intelligent Systems (2019).

Human-machine collaboration should be optimised to safeguard against poor decision-making including automation bias and/or mistrust of the system (Hoffman et al. 2018; Alexander 2019). AI should provide confidence and uncertainty in the information or choices being offered by an AI (Christensen and Lyons 2017; McLellan 2016).

*Trust – how can AI be trusted?*

Human-AI systems in Defence need to be trusted by users and operators, by commanders and support staff and by the military, government, and civilian population of a nation. MEAID points out the *High-Level Expert Group on Artificial Intelligence of the European Union* "believe it is essential that trust remains the bedrock of societies, communities, economies and sustainable development" (High-Level Expert Group on Artificial Intelligence 2019). They argue that trustworthy AI must be lawful, ethical, and robust.

MEAID suggests that trust is a relation between human-human, human-machine and machine-machine, consisting of two components: competency and integrity. Competence comprises of skills, reliability and experience; Integrity comprises of motives, honesty and character (Devitt 2018). This framework is consistent with the emphasis on character and professional competence in ADF-P-0 ADF Leadership Doctrine (Australian Defence Force 2021a, 4). It is noted that the third value of ADF leadership is *understanding*, which falls within the responsibility facet discussed above.

Operators will hold multiple levels of trust in the systems they are using depending on what aspect of trust is under scrutiny. In some cases, users may develop a reliance on low integrity technology that they can predict easily, such as using the known flight path of an adversary's drone to develop countermeasures. Users may also depend on technologies because of convenience rather than trust. Finally individual differences exist in the propensity to trust, highlighting that trust is a relational rather than an objective property.

To be trusted, AI systems need to be safe and secure within the Nation's sovereign supply chain. Throughout their lifecycle, AI systems should reliably operate in accordance with their intended purpose (Department of Industry Innovation and Science 2019).

*Law: How can AI be used lawfully?*

AI developers should be cognisant of the legal obligations within their anticipated use of the technology. Law within a Defence context has specific ethical considerations that must be understood. International humanitarian law (IHL) (lex specialis) and international human rights law (lex generalis) were forged from ethical theories in just war theory jus ad bellum governing the resort to force, jus in bello regulating the conduct of parties engaged in lawful combat (Coates 2016), jus post bellum regarding obligations after combat. The legal frameworks that accompany Defence activities are human-centred, which should mean that AI compliance with them will produce more ethical outcomes (Liivoja and McCormack 2016).

Using AI to augment human decision-making could lead to better humanitarian outcomes. There are many policies and directives that may apply, some of which have the force of law. In military context, there will also typically be an extant set of rules called the *rules of engagement*, which among other things specify the conditions that must be met in order to fire upon a target.

Legal compliance may be able to be 'built into' AI algorithms, but this relies on legal rules being sufficiently unambiguous and well specified that they can be encoded as rules that a computer can interpret and meets stakeholder expectations. In practice, laws are not always that clear, even to humans. Laws can intentionally be created with ambiguity to provide flexibility. In addition, they can have many complicated conditions and have many interconnections to other laws. Further work is needed to clarify how AI can best enable abidance with applicable laws.

*Traceability: How are the actions of AI recorded?*

MEAID notes that there are legislative requirements for Defence to record its decision-making. However, the increasing use of AI within human-AI systems means the manner of records must be considered. Records can represent the systems involved, the causal chain of events, and the humans and AIs that were part of decisions.

MEAID suggests that information needs to be accessible and explanatory; the training and expertise of humans must be open to scrutiny; and the background theories and assumptions, training, test and evaluation process of AIs must be retained. Information on AI systems should be available and understandable by auditors. Just as some aspects of human decision-making can be inscrutable, some aspects of the decisions of AIs may remain opaque. Emerging transparency standards may guide best practise for Defence (Winfield et al. 2021).

When decisions lead to expected outcomes or positive outcomes, the factors that lead to those decisions may not come under scrutiny. However, when low likelihood and/or negative outcomes occur, organisations should be able to 'rewind' the decision process to understand what occurred and what lessons might be learned. Noting that decisions made under uncertainty will always have a chance of producing negative outcomes, even if the decision-making process is defensible and operators are acting appropriately.

No matter how an AI is deployed in Defence, its data, training, theoretical underpinning, decision-making models and actions should be recorded and auditable by the appropriate levels of government and, where appropriate, made available to the public.

**Method for Ethical AI in Defence**

MEAID recommends assessing ethical compliance from design to deployment, requiring repeated testing, prototyping, and reviewing for technological and ethical limitations. Developers already must produce risk documentation for technical issues. Similar documentation for ethical risks ensures developers identify, acknowledge and attempt to mitigate ethical risks early in the design process and throughout test and evaluation (Devitt et al. 2021).

MEAID closely aligns to the IEEE Standard Model Process for Addressing Ethical Concerns During System Design (IEEE 2021)—see Box 7.

---

**Box 7 IEEE 7000-2021 standard** (IEEE 2021) provides:

- a system engineering standard approach integrating human and social values into traditional systems engineering and design.
- processes for engineers to translate stakeholder values and ethical considerations into system requirements and design practices.
- a systematic, transparent, and traceable approach to address ethically- oriented regulatory obligations in the design of autonomous intelligent systems.

---

Australia has developed a practical methodology (Devitt et al. 2021) that can support AI project managers and teams to manage ethical risks in military AI projects including three tools:

1. An AI Checklist for the development of ethical AI systems
2. An Ethical AI Risk Matrix to describe identified risks and proposed treatment

3. For larger programs, a data item descriptor (DID) for contractors to develop a formal Legal, Ethical and Assurance Program Plan (LEAPP) to be included in project documentation for AI programs where an ethical risk assessment is above a certain threshold (See APPENDIX G. DATA ITEM DESCRIPTION DID-ENG-SW-LEAPP Devitt et al. 2021)).

**AI Checklist**

The main components of the checklist are:

    A. Describe the military context in which the AI will be employed

    B. Explain the types of decisions supported by the AI

    C. Explain how the AI integrates with human operators to ensure effectiveness and ethical decision making in the anticipated context of use and countermeasures to protect against potential misuse

    D. Explain framework/s to be used

    E. Employ subject matter experts to guide AI development

    F. Employ appropriate verification and validation techniques to reduce risk.

Ethical AI Risk Matrix

An Ethical AI Risk Matrix will:

- Define the activity being undertaken
- Indicate the ethical facet and topic the activity is intended to address.
- Estimate the risk to the project objectives if issue is not addressed?
- Define specific actions you will undertake to support the activity
- Provide a timeline for the activity
- Define action and activity outcomes
- Identify the responsible party(ies)

- Provide the status of the activity.

**Analysis**

MEAID offers practical advice and tools for defence industries and Defence to communicate, document and iterate design specifications for emerging technologies and to identify operational contexts of use considerate of ethical and legal considerations and obligations. MEAID also offers entry points to explain system function, capability, and limits to both expert and non-expert stakeholders to military technologies.

MEAID aims to practically ensure accountability for a) considering ethical risks, b) assigning person(s) to each risk and c) making humans accountable for decisions on how ethics are de-risked. It has been noted on the International Committee of the Red Cross blog (Copeland and Sanders 2021) as establishing an iterative process to engage industry during the design and acquisition phase of new technologies to increase IHL abidance and reduce civilian harms.

Developed by Defence Science and Technology Group, Plan Jericho Air Force (Department of Defence 2020b) and Trusted Autonomous Systems[11]; the MEAID framework has been adopted by industry[12], is the ethics framework used in a case study of Allied Impact (Gaetjens, Devitt, and Shanahan 2021) and being trialled by Australia in the TTCP AI Strategic Challenge (Stanley-Lockman 2021, 43-44).

A side-by-side comparison between AU-EP, MEAID and OECD shows significant overlap in responsibility and trust, but also gaps where military uses of AI encounter ethical

---

[11] See https://tasdcrc.com.au/
[12] See Athena AI at https://athenadefence.ai/software

considerations not applicable to the civilian realm, such as the application of just war principles of distinction and proportionality (Law); and military control of weapons systems (Governance)—see ANNEX B. The AU-EP share many similarities with the OECD. For example, contestability is equivalent to OECD requirement that humans can understand and intervene on AI-based outcomes as well as challenge them

While not a formally adopted view of the Australian government, MEIAD establishes tools to assess ethical compliance that, "[e]ven as an opinion, the Method is the clearest articulation of ethical AI for defense among the Indo-Pacific allies" (Stanley-Lockman 2021, 21). As Stanley-Lockman (2021) states:

> "The [MEAID] tools offer a process to validate that contractors have indeed taken the ethical risks they identified into account in their design and testing prior to later acquisition phases….The incorporation of ethics in design through the acquisition lifecycle also intends to build trust in the process and, *by extension, the systems by the time they go into service.*"

Like many countries, ADO undertakes a formal capability acquisition process to assist the Government to identify and meet its military capability needs. This process, known as the Product Life Cycle, consist of four phases (strategy and concepts; risk mitigation and requirement setting; acquisition; and in-service and disposal) which are separated by Government decisions gates (Commonwealth of Australia 2020). This process ensures that Government's strategic objectives (see Box 1) drive Defence's acquisition priorities.

Australia's Sovereign Industry Capability Priorities (see Box 2) reflects the Government's realisation that future Defence AI capabilities will increasingly rely on research and development in the civil sector. This necessitates closer collaboration between Defence and Defence industry to ensure the timely delivery of cutting-edge technology the reflects Australia's values (see Box 3) and ethical AI principles (see Box 4) and ensures legal and ethical risks associated with military AI technology are identified and mitigated in the earliest stages of development.

Australia's Department of Defence may inform Defence industry its legal and ethical requirements to enable AI developers to introduce design measures to mitigate or remove the risks before entering the Defence procurement process, rather than attempting to address legal and ethical risks during the acquisition process. This provides efficacies for both Defence and industry by allowing industry to better focus their development priorities and assists Defence in streamlining its AI capability acquisition process.

MEAID provides Defence with a practical approach that can readily integrate into the existing Product Life Cycle process to inform and enable the transfer of legal and ethical AI technology from Defence industry into Defence. MEAID tools such as the LEAPP provide Defence with visibility of a contractor's plan to mitigate legal and ethical risk and, together with the facets of ethical AI (see Figure 2) and the Article 36 weapon review process, can inform Government decisions to acquire military AI technology that both legal and align with Australia's ethical principles.

Australia can play a leadership role by integrating legal and ethical considerations into its Defence AI capability acquisition process. This requires a policy framework that defines its legal and ethical requirements, is informed by Defence industry stakeholders and provides a practical methodology to integrate legal and ethical risk mitigation strategies into the acquisition process.

**Conclusion**

This chapter explored Australia's public positioning on AI and AI governance 2018-2021 through published strategies, frameworks, action plans and government reports. While these provide a top-down view, high-level national AI strategies may align with the lived experience of public servants and personnel encountering AI in Defence or Australian Defence Force commanders and operators using AI systems (Kuziemski and Misuraca 2020).

Australia is a leading AI nation with strong allies and partnerships. It has prioritised the development of robotics, AI, and autonomous systems to develop sovereign capability for the military. Australia commits to Article 36 reviews of all new means and method of warfare to ensure weapons and weapons systems are operated within acceptable systems of control. Additionally, the country has undergone significant reviews of the risks of AI to human rights and within intelligence organisations and has committed to producing ethics guidelines and frameworks in Security and Defence (Department of Defence 2021a; Attorney-General's Department 2020). Australia is committed to OECD's values-based principles for the responsible stewardship of trustworthy AI as well as adopting a set of National AI ethics principles. While Australia has not adopted an AI governance framework specifically for Defence; *A Method for Ethical AI in Defence* (MEAID) published by Defence Science includes a framework and pragmatic tools for managing ethical and legal risks for military applications of AI.

Key findings of the chapter are that Australia has formed strong international AI governance partnerships likely to reinforce and strengthen strategic partnerships and power relations. Like many nations, Australia's commitment to civilian AI Ethics principles do not provide military guidance or governance. The ADO has the opportunity to adopt a robust AI ethical policy for security and defence that emphasises commitment to existing international legal frameworks and can be applied to AI-driven weapons. A risk-based ethical AI framework suited for military purposes and aligned with best practise, standards and frameworks internationally can ensure defence industries consider ethics-by-design and law-by-design ahead of the acquisitions process. Australia should continue to invest, research and develop AI governance frameworks to meet the technical potential and strategic requirements of military uses of AI.

**Bibliography**

88th Air Base Wing Public Affairs, 23 December, 2019, "Digital engineering transformation coming to Air Force weapons enterprise," https://www.af.mil/News/Article-Display/Article/2046599/digital-engineering-transformation-coming-to-air-force-weapons-enterprise/.

AI Council. 2021. "AI Roadmap." United Kingdom. https://www.gov.uk/government/publications/ai-roadmap.

Alexander, Donovan. 2019. "Is our reliance on technology creating a new dark age?" *Interesting Engineering*, 10 May, 2019. https://interestingengineering.com/is-our-reliance-on-technology-creating-a-new-dark-age.

Army. 2018. "Robotic and Autonomous Systems Strategy." https://researchcentre.army.gov.au/library/other/robotic-autonomous-systems-strategy.

ASPI, "Russia–Ukraine war, policing and AI, and an Australian DARPA," 4 March, 2022, in *Policy, Guns and Money*, https://www.aspistrategist.org.au/policy-guns-and-money-russia-ukraine-war-policing-and-ai-and-an-australian-darpa/.

Attorney-General's Department. 2020. "Government response to the Comprehensive review of the legal framework of the National Intelligence Community." Accessed 25 September 2021. https://www.ag.gov.au/national-security/publications/government-response-comprehensive-review-legal-framework-national-intelligence-community.

Australian Defence Doctrine Publication. 2021. ADF-P-7 Learning.

Australian Defence Force. 2021a. "ADF-P-0 ADF Leadership, Edition 3." The Forge. https://theforge.defence.gov.au/adf-philosophical-doctrine-adf-leadership

---. 2021b. "ADF-P-0 Military Ethics, Edition 1, 2021." Accessed 15 October. https://theforge.defence.gov.au/ethics.

Australian Government. 2019. "Australia's System of Control and applications for Autonomous Weapon Systems." Group of Governmental Experts on Emerging Technologies in the Area of Lethal Autonomous Weapons Systems, Geneva, 25–29 March 2019 and 20–21 August 2019. Accessed 10 September. https://docs-library.unoda.org/Convention_on_Certain_Conventional_Weapons_-_Group_of_Governmental_Experts_(2019)/CCWGGE.12019WP.2Rev.1.pdf.

---. 2021. "Four new Sovereign Industrial Capability Priorities announced." 7 September, 2021. https://business.gov.au/cdic/news-for-defence-industry/four-new-sovereign-industrial-capability-priorities-announced.

Australian Permanent Mission and Consulate-General Geneva. 2017. "Australian Statement - General Exchange of Views, LAWS GGE 13-17 November 2017." https://geneva.mission.gov.au/gene/Statement783.html.

Australian Research Council. 2020. "Codes and Guidelines." https://www.arc.gov.au/policies-strategies/policy/codes-and-guidelines.

Australian Signals Directorate. 2019a. "ASD Corporate Plan 2019-2020." https://www.asd.gov.au/sites/default/files/2019-08/ASD_Corporate_Plan_final_12.pdf.

---. 2019b. "Values." Accessed 25 September. https://www.asd.gov.au/about/values.

Blades, Johnny. 2021. "Aukus pact strikes at heart of Pacific regionalism." *Radio New Zealand Pacific*, 2021. https://www.rnz.co.nz/international/pacific-news/451715/aukus-pact-strikes-at-heart-of-pacific-regionalism.

Braithwaite, Valerie. 2020. "Beyond the bubble that is Robodebt: How governments that lose integrity threaten democracy." *Australian Journal of Social Issues* 55 (3): 242-259.

Burgess, M. 2019. "What ASD cyber operatives really do to protect Australian interests." 28 March, 2019. https://www.themandarin.com.au/106332-mike-burgess-director-general-asd-speech-to-the-lowy-institute/.

Centre for Work Health and Safety. 2021. "Ethical use of artificial intelligence in the workplace - AI WHS Scorecard." NSW Government. https://www.centreforwhs.nsw.gov.au/knowledge-hub/ethical-use-of-artificial-intelligence-in-the-workplace-final-report.

Christensen, James C., and Joseph B. Lyons. 2017. "Trust between Humans and Learning Machines: Developing the Gray Box." *Mechanical Engineering* 139 (06): S9-S13. https://doi.org/10.1115/1.2017-Jun-5. https://doi.org/10.1115/1.2017-Jun-5.

Cihon, Peter. 2019. "Standards for AI governance: international standards to enable global coordination in AI research & development." *Future of Humanity Institute. University of Oxford*. https://www.fhi.ox.ac.uk/wp-content/uploads/Standards_-FHI-Technical-Report.pdf

Commonwealth of Australia. 2018. "The Australian Article 36 Review Process." United Nations Group of Governmental Experts of the High Contracting Parties to the Convention on Prohibitions or Restrictions on the Use of Certain Conventional Weapons Which May Be Deemed to Be Excessively Injurious or to Have Indiscriminate Effects, 30 August 2018. https://docs-library.unoda.org/Convention_on_Certain_Conventional_Weapons_-_Group_of_Governmental_Experts_(2018)/2018_GGE%2BLAWS_August_Working%2Bpaper_Australia.pdf.

---. 2019. "Australia's System of Control and applications for Autonomous Weapon Systems." Group of Governmental Experts on Emerging Technologies in the Area of Lethal Autonomous Weapons Systems, 25–29 March 2019 and 20–21 August 2019, Geneva, 26 March 2019. https://www.unog.ch/80256EDD006B8954/(httpAssets)/16C9F75124654510C12583C9003A4EBF/$file/CCWGGE.12019WP.2Rev.1.pdf.

---. 2020. Capability Life Cycle Manual (V.2.1). edited by Investment Portfolio Management Branch.

Copeland, D., and L. Sanders. 2021. "Engaging with the industry: integrating IHL into new technologies in urban warfare." *Humanitarian Law and Policy* (blog), *ICRC*. 8 October. https://blogs.icrc.org/law-and-policy/2021/10/07/industry-ihl-new-technologies/.

Cryer, Robert, Darryl Robinson, and Sergey Vasiliev. 2019. *An introduction to international criminal law and procedure*. Cambridge University Press.

Cyber Digital and Technology Policy Division. 2020. "2020 Cyber Security Strategy." Department of Home Affairs. https://www.homeaffairs.gov.au/about-us/our-portfolios/cyber-security/strategy.

Dafoe, Allan. 2018. "AI governance: a research agenda." *Governance of AI Program, Future of Humanity Institute, University of Oxford: Oxford, UK* 1442: 1443.

Dawson, D., E. Schleiger, J. Horton, J. McLaughlin, C. Robinson, G. Quezada, J. Scowcroft, and S. Hajkowicz. 2019. *Artificial Intelligence: Australia's Ethics Framework: A Discussion Paper.* Data61 CSIRO, Australia (Data61 CSIRO, Australia: Australia Data61 CSIRO). https://consult.industry.gov.au/strategic-policy/artificial-intelligence-ethics-framework/.

de Git, Melanie 2021. "Loyal Wingman uncrewed aircraft completes first flight." *Innovation Quarterly, Boeing*, 12 April, 2021. https://www.boeing.com/features/innovation-quarterly/2021/04/loyal-wingman.page.

Defence Science & Technology Group, 18 November, 2021a, "AI to enable military commanders to make better decisions," https://www.dst.defence.gov.au/news/2021/11/18/ai-enable-military-commanders-make-better-decisions-faster.

---. 2021b. "Defence Artificial Intelligence Research Network (DAIRNET) Research Call." Accessed 1 November. https://www.dst.defence.gov.au/partner-with-us/opportunities/defence-artificial-intelligence-research-network-dairnet-research-call.

Defence Science Institute, 19 May, 2020a, "The Artificial Intelligence for Decision Making Initiative," https://www.defencescienceinstitute.com/news/the-artificial-intelligence-for-decision-making-initiative.

---, 2020b, "'Artificial Intelligence for Decision Making' Initiative," https://www.defencescienceinstitute.com/component/sppagebuilder/?view=page&id=29&highlight=WyJhcnRpZmljaWFsIiwiJ2FydGlmaWNpYWwiLCJpbnRlbGxpZ2VuY2UiLCJhcnRpZmljaWFsIGludGVsbGlnZW5jZSJd.

---, 10 May, 2021, "Applications open for the artificial intelligence for decision making initiative round 2," https://www.defencescienceinstitute.com/news/initiatives/applications-open-for-the-artificial-intelligence-for-decision-making-initiative-round-2.

Deloitte Center for Government Insights. 2021. "The future of warfighting." Deloitte. https://www2.deloitte.com/global/en/pages/public-sector/articles/future-of-warfighting.html.

Department of Defence. 2016. "Defence White Paper." https://www1.defence.gov.au/about/publications/2016-defence-white-paper.

---. 2019. ADDP 00.1 Command and Control AL1. edited by Department of Defence.

---. 2020a. "2020 Defence Strategic Update." https://www1.defence.gov.au/about/publications/2020-defence-strategic-update.

---. 2020b. "Artificial intelligence enhances the impact of air and space power for the Joint Force." Department of Defence Annual Report 2019-2020. https://www.transparency.gov.au/annual-reports/department-defence/reporting-year/2019-20-31.

---. 2020c. "Lead the Way: Defence Transformation Strategy." https://www1.defence.gov.au/about/publications/lead-way-defence-transformation-strategy.

---. 2021a. "Defence Data Strategy 2021-2023." https://www1.defence.gov.au/about/publications/defence-data-strategy-2021-2023.

---, 16 February, 2021b, "Defence releases report on ethical use of AI," https://news.defence.gov.au/media/media-releases/defence-releases-report-ethical-use-ai.

Department of Defense. 1978. Managing Weapon System Software: Progress and Problems (Unclassified Digest of a Classified Report).

Department of Home Affairs. 2020. "Australian Values." https://www.homeaffairs.gov.au/about-us/our-portfolios/social-cohesion/australian-values.

Department of Industry Innovation and Science. 2019. "Australia's Artificial Intelligence Ethics Framework." Accessed 25 September. https://www.industry.gov.au/data-and-publications/australias-artificial-intelligence-ethics-framework/australias-ai-ethics-principles.

Department of Industry Science Energy and Resources, 16 June 2020, 2020a, "The Global Partnership on Artificial Intelligence launches,"

- https://www.industry.gov.au/news/the-global-partnership-on-artificial-intelligence-launches.
- ---. 2020b. "Testing the AI Ethics Principles." Accessed 11 December https://www.industry.gov.au/data-and-publications/australias-artificial-intelligence-ethics-framework/testing-the-ai-ethics-principles.
- ---. 2021. "Australia's Artificial Intelligence Action Plan." https://www.industry.gov.au/data-and-publications/australias-artificial-intelligence-action-plan.
- Devitt, S K 2018. "Trustworthiness of autonomous systems." In *Foundations of Trusted Autonomy*, edited by Hussein A. Abbass, Jason Scholz and Darryn J. Reid, 161-184. Cham: Springer International Publishing.
- Devitt, S K, M Gan, J Scholz, and R S Bolia. 2021. *A Method for Ethical AI in Defence.* Defence Science and Technology Group (Defence Science and Technology). https://www.dst.defence.gov.au/publication/ethical-ai.
- Fisher, Erik. 2020. "Necessary conditions for responsible innovation." *Journal of Responsible Innovation* 7 (2): 145-148. https://doi.org/10.1080/23299460.2020.1774105. https://doi.org/10.1080/23299460.2020.1774105.
- Gaetjens, D., S.K. Devitt, and C. Shanahan. 2021. *Ethical AI in Defence Case Study: Allied Impact. DST Technical Report*: Defence Science & Technology Group.
- Galloway, Kate. 2017. "Big Data: A case study of disruption and government power." *Alternative Law Journal* 42 (2): 89-95.
- Gasser, Urs, and Virgilio AF Almeida. 2017. "A layered model for AI governance." *IEEE Internet Computing* 21 (6): 58-62.
- Gobal Partnership on AI. 2021. "The Global Partnership on AI ". Accessed 25 September https://gpai.ai/.
- Hajkowicz, S A, S Karimi, T Wark, C Chen, M Evans, N Rens, D Dawson, A Charlton, T Brennan, C Moffatt, S Srikumar, and K J Tong. 2019. *Artificial Intelligence: Solving problems, growing the economy and improving our quality of life.* (CSIRO Data61 and the Department of Industry, Innovation and Science, Australian Government). https://data61.csiro.au/en/Our-Research/Our-Work/AI-Roadmap.
- Hanson, Fergus, and Danielle Cave. 2021. "Australia well placed to turbocharge its strategic tech capability." Australian Strategic Policy Institute. Last Modified 20 September. https://www.aspi.org.au/opinion/australia-well-placed-turbocharge-its-strategic-tech-capability.
- High-Level Expert Group on Artificial Intelligence. 2019. *Ethics Guidelines for Trustworthy AI.* European Commission. https://ec.europa.eu/digital-single-market/en/news/ethics-guidelines-trustworthy-ai.
- Hoffman, Robert R., Nadine Sarter, Matthew Johnson, and John K. Hawley. 2018. "Myths of automation and their implications for military procurement." *Bulletin of the Atomic Scientists* 74 (4): 255-261. https://doi.org/10.1080/00963402.2018.1486615.
- IEEE. 2021. "IEEE 7000™-2021 - IEEE Standard Model Process for Addressing Ethical Concerns During System Design." https://engagestandards.ieee.org/ieee-7000-2021-for-systems-design-ethical-concerns.html.
- IEEE Global Initiative on Ethics of Autonomous and Intelligent Systems. 2019. *Ethically Aligned Design: A Vision for Prioritizing Human Well-being with Autonomous and Intelligent Systems (EADe1).* IEEE. https://standards.ieee.org/content/ieee-standards/en/industry-connections/ec/autonomous-systems.html.
- Insinna, Valerie. 2021. "Australia makes another order for Boeing's Loyal Wingman drones after a successful first flight." *DefenceNews*, 3 March, 2021. https://www.defensenews.com/air/2021/03/02/australia-makes-another-order-for-boeing-made-loyal-wingman-drones-after-a-successful-first-flight/.

International Committee of the Red Cross. 2006. "A Guide to the Legal Review of New Weapons, Means and Methods of Warfare: Measures to Implement Article 36 of Additional Protocol I of 1977." *International Review of the Red Cross* 88 (864). https://www.icrc.org/eng/assets/files/other/irrc_864_icrc_geneva.pdf.

JAIC Public Affairs, 16 September 2020, 2020, "JAIC facilitates first-ever International AI Dialogue for Defense," https://www.ai.mil/news_09_16_20-jaic_facilitates_first-ever_international_ai_dialogue_for_defense_.html.

---, 28 May, 2021, "DoD Joint AI Center Facilitates Third International AI Dialogue for Defense," https://www.ai.mil/news_05_28_21-jaic_facilitates_third_international_ai_dialogue_for_defense.html.

Jenkins, K. 2014. "Collaboration for cultural reform in Defence." https://defence.humanrights.gov.au/.

Kelly, Paul. 2022. "New world disorder: Ukraine redefines global landscape." 4 March, 2022. https://www.theaustralian.com.au/inquirer/a-new-world-disorder-morrison-calls-on-the-west-to-unite-against-russia-and-china/news-story/433b6ff74637c45454393030f827f1e7.

Kuziemski, Maciej, and Gianluca Misuraca. 2020. "AI governance in the public sector: Three tales from the frontiers of automated decision-making in democratic settings." *Telecommunications policy* 44 (6): 101976.

Liivoja, Rain, Eve Massingham, Tim McFarland, and Simon McKenzie. 2020. "Are Autonomous Weapons Systems Prohibited?". Game Changer. Trusted Autonomous Systems https://tasdcrc.com.au/are-autonomous-weapons-systems-prohibited/.

Lockey, S., N. Gillespie, and C. Curtis. 2020. *Trust in artificial intelligence: Australian insights 2020.* (The University of Queensland and KPMG Australia). https://assets.kpmg/content/dam/kpmg/au/pdf/2020/public-trust-in-ai.pdf.

Lopez, C.T. 2020. "DOD Adopts 5 Principles of Artificial Intelligence Ethics." *DOD News*, 2020. https://www.defense.gov/Explore/News/Article/Article/2094085/dod-adopts-5-principles-of-artificial-intelligence-ethics/.

Lowy Institute. 2019. "Mike Burgess, Director-General on the Australian Signals Directorate (ASD) – Offensive cyber" YouTube. https://youtu.be/Th6EKCwhGrs.

McFarland, Tim. 2020. *Autonomous Weapon Systems and the Law of Armed Conflict: Compatibility with International Humanitarian Law*. Cambridge Core.

---. 2021. "Autonomous Weapons and The Jus Ad Bellum." *Law School Policy Review*. https://lawschoolpolicyreview.com/2021/03/20/autonomous-weapons-and-the-jus-ad-bellum-an-overview/.

McLellan, Charles. 2016. "Inside the black box: Understanding AI decision-making." *ZDNet*, 1 December, 2016. https://www.zdnet.com/article/inside-the-black-box-understanding-ai-decision-making/.

Ministère des Armées. 2019. Artificial Intelligence in Support of Defence: Report of the AI Task Force.

Morrison, B. 2021. "How will artificial intelligence and machine learning impact OHS?". Australian Institute for Health and Safety. Last Modified 21 June. https://www.aihs.org.au/news-and-publications/news/how-will-artificial-intelligence-and-machine-learning-impact-ohs.

Morrison, S, B Johnson, and J Biden. 2021. "Remarks by President Biden, Prime Minister Morrison of Australia, and Prime Minister Johnson of the United Kingdom Announcing the Creation of AUKUS." The White House. https://www.whitehouse.gov/briefing-room/speeches-remarks/2021/09/15/remarks-by-president-biden-prime-minister-morrison-of-australia-and-prime-minister-johnson-of-the-united-kingdom-announcing-the-creation-of-aukus/.

National Security Commission on Artificial Intelligence. 2021. "Final Report." United States of America. https://www.nscai.gov/wp-content/uploads/2021/03/Full-Report-Digital-1.pdf.

Nicholson, B. 2021. "Morrison says AUKUS will strengthen cooperation on critical technologies." *The Strategist, Australian Strategic Policy Institute*, 17 November, 2021. https://www.aspistrategist.org.au/morrison-says-aukus-will-strengthen-cooperation-on-critical-technologies/.

NSW Government. 2022. "NSW AI Assurance Framework." NSW Government. Accessed 5 March. https://www.digital.nsw.gov.au/policy/artificial-intelligence/nsw-ai-assurance-framework.

OECD. 2019. "The OECD AI Principles." https://www.oecd.org/going-digital/ai/principles/.

OECD Council on Artificial Intelligence. 2019. Recommendation of the Council on Artificial Intelligence.

Office of the Director of National Intelligence. "Five Eyes Intelligence Oversight and Review Council (FIORC)." Accessed 25 September. https://www.dni.gov/index.php/ncsc-how-we-work/217-about/organization/icig-pages/2660-icig-fiorc.

Parikh, Nish. 2021. "Understanding Bias In AI-Enabled Hiring." *Forbes Magazine*, 2021. https://www.forbes.com/sites/forbeshumanresourcescouncil/2021/10/14/understanding-bias-in-ai-enabled-hiring/?sh=33f2d7997b96.

Persley, Alexandra, 2021, "Australia claims historic top two spot in the 'Robot Olympics'," https://www.csiro.au/en/news/News-releases/2021/Australia-claims-historic-top-two-spot-in-the-Robot-Olympics.

Prime Minister, and Minister of Defence. 2022, 1 March 2022. "Australian Support to the Ukraine." Prime Minister of Australia. Accessed 5 March 2022. https://www.pm.gov.au/media/australian-support-ukraine.

Richardson, D. 2020. "Report of the comprehensive review of the legal framework of the national intelligence community." Attorney-General's Department. Accessed 25 September. https://www.ag.gov.au/national-security/consultations/comprehensive-review-legal-framework-governing-national-intelligence-community.

Royal Australian Air Force. 2019. "At the edge: Exploring and exploiting our fifth-generation edges." https://www.airforce.gov.au/our-mission/plan-jericho.

---. 2021. "Loyal Wingman First Flight." *YouTube*, 2 March, 2021. https://youtu.be/BiSHVl7UMRk.

Royal Australian Navy. 2020. "RAS-AI Strategy 2040: Warfare Innovation Navy." https://www.navy.gov.au/media-room/publications/ras-ai-strategy-2040.

Sadler, Denham. 2021. "HRC calls for an AI Safety Commissioner." *InnovationAus*, 27 May, 2021. https://www.innovationaus.com/hrc-calls-for-an-ai-safety-commissioner/.

Santow, E. 2021. "Human Rights and Technology Final Report." Australian Human Rights Commission. https://tech.humanrights.gov.au/downloads.

Schmitt, Lewin. 2021. "Mapping global AI governance: a nascent regime in a fragmented landscape." *AI and Ethics*. https://doi.org/10.1007/s43681-021-00083-y. https://doi.org/10.1007/s43681-021-00083-y.

Selwyn, Neil, and Beatriz Gallo Cordoba. 2021. "Australian public understandings of artificial intelligence." *AI & SOCIETY*: 1-18. https://link.springer.com/article/10.1007/s00146-021-01268-z.

Senate Foreign Affairs Defence and Trade Legislation Committee. 2019. "Official Committee Hansard Senate Foreign Affairs, Defence and Trade Legislation Committee Estimates Wednesday, 23 October 2019." https://parlinfo.aph.gov.au/parlInfo/search/display/display.w3p;query=Id%3A%22committees%2Festimate%2F53068544-efe7-4494-a0f2-2dbca4d2607b%2F0000%22.

Shih, Gerry, and Anne Gearan. 2021. "As Biden hosts first Quad summit at the White House, China is the background music." *The Washington Post*, 24 September, 2021. https://www.washingtonpost.com/world/2021/09/24/quad-us-india-australia-japan-china/.

Standards Australia. 2020. *Artificial Intelligence Standards Roadmap: Making Australia's Voice Heard.* https://www.standards.org.au/news/standards-australia-sets-priorities-for-artificial-intelligence.

Stanley-Lockman, Z. 2021. "Responsible and Ethical Military AI Allies and Allied Perspectives: CSET Issue Brief." Centre for Security and Emerging Technology, Georgetown University's Walsh School of Foreign Service. https://cset.georgetown.edu/wp-content/uploads/CSET-Responsible-and-Ethical-Military-AI.pdf.

The White House. 2021a. "FACT SHEET: U.S.-EU Establish Common Principles to Update the Rules for the 21st Century Economy at Inaugural Trade and Technology Council Meeting." https://www.whitehouse.gov/briefing-room/statements-releases/2021/09/29/fact-sheet-u-s-eu-establish-common-principles-to-update-the-rules-for-the-21st-century-economy-at-inaugural-trade-and-technology-council-meeting/.

---. 2021b. "Joint Leaders Statement on AUKUS." Last Modified 15 September 2021. https://www.whitehouse.gov/briefing-room/statements-releases/2021/09/15/joint-leaders-statement-on-aukus/.

Thi Ha, Hoang 2021. "The Aukus challenge to Asean." *The Straits Times*, 2021. https://www.straitstimes.com/opinion/the-aukus-challenge-to-asean.

Townshend, Ashley, Thomas Lonergan, and Toby Warden. 2021. "The U.S.-Australian Alliance Needs a Strategy to Deter China's Gray-Zone Coercion." *War on the Rocks*, 2021. https://warontherocks.com/2021/09/the-u-s-australian-alliance-needs-a-strategy-to-deter-chinas-gray-zone-coercion/.

van Noorden, Richard. 2020. "The ethical questions that haunt facial-recognition research." *Nature*, 18 November, 2020. https://www.nature.com/articles/d41586-020-03187-3.

Vine, R. 2020. *Concept for robotics and autonomous systems.* Australian Defence Force. https://www.defence.gov.au/vcdf/forceexploration/adf-concept-future-robotics-autonomous-systems.asp.

White, L. 2021. "Tackling the growing threats to Australia's cyber security." *2021 Mandarin Defence Special Report*, 2021. https://www.themandarin.com.au/169281-tackling-the-growing-threats-to-australias-cyber-security/.

Winfield, Alan F. T., Serena Booth, Louise A. Dennis, Takashi Egawa, Helen Hastie, Naomi Jacobs, Roderick I. Muttram, Joanna I. Olszewska, Fahimeh Rajabiyazdi, Andreas Theodorou, Mark A. Underwood, Robert H. Wortham, and Eleanor Watson. 2021. "IEEE P7001: A Proposed Standard on Transparency." *Frontiers in Robotics and AI* 8 (225). https://doi.org/10.3389/frobt.2021.665729. https://www.frontiersin.org/article/10.3389/frobt.2021.665729.

Combat/Warfighting

| Tag | **Force Application (FA)** |
| --- | --- |
| Description | The conduct of military missions to achieve decisive effects through kinetic and non-kinetic offensive means. |
| AI examples | Autonomous weapons (AWs) and autonomous/semi-autonomous combat vehicles and subsystems |
| | AI used to support strategic, operational and tactical planning, including optimisation and deployment of major systems |
| | AI used in modelling and simulation used for planning and mission rehearsal |
| | AI used in support of the targeting cycle including for collateral damage estimation |
| | AI used for Information Warfare such as a Generative Adversarial Network (GAN-) generated announcement or strategic communication |
| | AI used to identify potential vulnerabilities in an adversary force to attack |
| | AI used for discrimination of combatants and non-combatants |

| Tag | **Force Protection (FP)** |
|---|---|
| **Description** | All measures to counter threats and hazards to, and to minimise vulnerabilities of, the joint force in order to preserve freedom of action and operational effectiveness |
| **AI examples** | Autonomous defensive systems (i.e. Close in Weapons Systems) |
| | AI used for Cyber Network Defence |
| | AI used to develop and employ camouflage and defensive deception systems and techniques |
| | Autonomous decoys and physical, electro-optic or radio frequency countermeasures |
| | AI to identify potential vulnerabilities in a friendly force that requires protection |
| | AI used to simulate potential threats for modelling and simulation or rehearsal activities |
| | Autonomous Medical Evacuation/Joint Personnel Recovery systems |

| Tag | **Force Sustainment (FS)** |
|---|---|
| **Description** | Activities conducted to sustain fielded forces, and to establish and maintain expeditionary bases. Force sustainment includes the provision of personnel, logistic and any other form of support required to maintain and prolong operations until accomplishment of the mission. |
| **AI examples** | Autonomous combat logistics and resupply vehicles |
| | Automated combat inventory management |

Predictive algorithms for the expenditure of resources such as fuel, spares and munitions

Medical AI systems used in combat environments and expeditionary bases

Predictive algorithms for casualty rates for personnel and equipment

Algorithms to optimise supply chains and the recovery, repair and maintenance of equipment

Algorithms to support the provision of information on climate, environment and topography

AI used for battle damage repair and front-line maintenance

| Tag | Situational Understanding (SU) |
|---|---|
| Description | The accurate interpretation of a situation and the likely actions of groups and individuals within it. Situational Understanding enables timely and accurate decision making. |
| AI examples | AI that enables or supports Intelligence, Surveillance and Reconnaissance (ISR) activities including: |
| | object recognition and categorisation of still and full motion video |
| | removal of unwanted sensor data |
| | identification of enemy deception activities |
| | anomaly detection and alerts |
| | monitoring of social media and other open-source media channels |
| | optimisation of collection assets |
| | AI that fuses data and disseminates intelligence to strategic, operational and tactical decision makers |
| | Decision support tools |
| | Battle Management Systems |
| | AI that supports Command and Control functions |
| | Algorithms used to predict likely actions of groups and individuals |
| | AI used to assess individual and collective behaviour and attitudes |

Enterprise-level and Rear Echelon Functions

| Tag | Personnel (PR) |
|---|---|
| Description | All activities that support the Raising, Training and Sustaining (RTS) of personnel. |
| AI examples | AI used for **Human Resource Management** including: |

record keeping

posting and promotion

disciplinary and performance management

recruitment and retention

modelling of future personnel requirements

prediction of HR supply and demand events and anomalies

AI used in individual and collective **training and education** including modelling and simulation

AI used for testing and certification of personnel

AI used to model the capability and preparedness of permanent and reserve personnel

| Tag | **Enterprise Logistics (EL)** |
|---|---|
| **Description** | Activities that support rear-echelon enterprise-level logistics functions including support of permanent military facilities |
| **AI examples** | Autonomous rear-echelon supply vehicles and warehouses |
| | AI used for optimisation of rear-echelon supply chains and inventory management |
| | AI used in depot-level and intermediate maintenance, including: |
| | Digital twinning |
| | Predictive maintenance |
| | Global supply chain analysis, prediction and optimisation |
| | Enterprise-level analysis and prediction for resource demand and supply (i.e. national/strategic fuel requirements) |

| Tag | **Business Process Improvement (BP)** |
|---|---|
| **Description** | Activities that support rear-echelon administrative business processes that are not related to personnel or logistics. |
| **AI examples** | AI used for Information Management and record-keeping |
| | Informational assistants such as policy chatbots |
| | AI that supports management of policy and procedures |
| | AI used to optimise business and administrative processes, including modelling and simulation tools |
| | AI used for enterprise business planning at the strategic, operational and tactical level |

ANNEX B Side-by-Side Comparison of AI Ethics Frameworks

| Facets of Ethical AI in Defence | Australian Government's AI Ethics Principles | OECD |
|---|---|---|
| **RESPONSIBILITY: Who is responsible for AI?** | Human, social and environmental wellbeing: Throughout their lifecycle, AI systems should benefit individuals, society and the environment<br><br>Human-centred values: Throughout their lifecycle, AI systems should respect human rights, diversity, and the autonomy of individuals | 1. AI should benefit people and the planet by driving inclusive growth, sustainable development and well-being.<br><br>2. AI systems should be designed in a way that respects … human rights, democratic values and diversity, |
| **GOVERNANCE: How is AI controlled?** | Accountability: Those responsible for the different phases of the AI system lifecycle should be identifiable and accountable for the outcomes of the AI systems, and human oversight of AI systems should be enabled | 5. Organisations and individuals developing, deploying or operating AI systems should be held accountable for their proper functioning in line with the above principles |
| | Transparency and explainability: There should be transparency and responsible disclosure to ensure people know when they are being significantly impacted by an AI system, and can find out when an AI system is engaging with them<br><br>Contestability: When an AI system significantly impacts a person, community, group or environment, there should be a timely process to allow people to challenge the use or output of the AI system | 3. There should be transparency and responsible disclosure around AI systems to ensure that people understand AI-based outcomes and can challenge them. |
| **TRUST:**<br>**How can AI be trusted?** | Reliability and safety: Throughout their lifecycle, AI systems should reliably operate in accordance with their intended purpose<br><br>Fairness: Throughout their lifecycle, AI systems should be | 2. [AI systems] should include appropriate safeguards – for example, enabling human intervention where necessary – to ensure a fair and just society. |
| | inclusive and accessible, and should not involve or result in unfair discrimination against individuals, communities or groups

Privacy protection and security: Throughout their lifecycle, AI systems should respect and uphold privacy rights and data protection, and ensure the security of data | 4. AI systems must function in a robust, secure and safe way throughout their life cycles and potential risks should be continually assessed and managed. |
| **LAW:** **How can AI be used lawfully?** | No equivalent | 2. AI systems should be designed in a way that respects the rule of law |
| **TRACEABLILITY:** **How are the actions of AI recorded?** | No equivalent (but implied) | No equivalent (but implied) |